\documentstyle[12pt, epsfig, aps,prb,preprint]{revtex}
\begin{document}

\title{Exact results for the one dimensional periodic Anderson model at 
finite $U$}

\author{Iv\'an Orlik and Zsolt Gul\'acsi}

\address{Department of Theoretical Physics, Lajos Kossuth University,\\
         H-4010 Debrecen, Poroszlay \'ut 6/c, P.O. Box 5, Hungary}

\date{February, 1999}

\maketitle

\begin{abstract}
We present  exact  results for  the periodic Anderson model
for finite Hubbard interaction $0 \:  \leq U \: < \: \infty$ on certain 
restricted domains of the model's phase diagram, in $d \: = \: 1$ dimension.
Decomposing the Hamiltonian into positive semidefinite terms we find two 
quantum states to be ground state, an insulating and a metallic one.
The ground state energy  and several ground state expectation values are 
calculated. 
\end{abstract}

\section{Introduction}
Since the last decades compounds with lanthanide or actinide elements 
as heavy-fermion systems have attracted great interest because of their 
interesting experimental behavior. Numerous strange phases are thought to be 
possible as the ground state of such a system, ordered  and disordered
phases as well. As a theoretical background  the periodic Anderson model (PAM)
is one of the most important  applicants to explain the physics in these 
systems. However, despite its relatively simple structure up to now the 
complete exact solution of the model is not known even in one dimension and 
there are mostly approximative results. In 1993, using a method proposed by 
Brandt and Giesekus \cite{b}, Strack \cite{s} found the exact ground state 
in $d \: = \: 1$ for the strong coupling $U \: = \: \infty$ case. His results
were generalized for $d \: = \: 2 \: , \: 3 \:$ dimensions by Orlik and 
Gul\'acsi \cite{o} also in $U \: = \: \infty$. This strong coupling limit is 
however an extremal case and is important to look for results in the physical
 domain  where 
$U\: < \: \infty$. Here we present for the first time exact results for PAM 
at finite $U \: < \: \infty$, in $d \: = \: 1$ dimension. Two different 
solutions are reported, one insulating and a metallic one, both at 3/4
filling. The paper is organized as follows: The following Section is devoted
to introduce the model and to show the positive decomposed form of the 
Hamiltonian used in the remaining part of the paper. Section III. contains 
the results of the calculations, and finally Section IV. concludes the 
results and closes the paper. 
 
\section{The Hamiltonian}

To start our study we use the Hamiltonian ($H$) of PAM as presented below:

\begin{eqnarray} \label{Hamiltonian}
H \: = \:  \hat{T}_{f} \: + \: \hat{T}_{c} \: + \: \hat{V}_{0}
\: + \: \hat{V}_{1} \: + \: E_{f} \: \sum_{i, \sigma} \: 
\hat{n}_{i, \sigma}^{f} \: + \: U \: \hat{U} \: ,
\end{eqnarray}

where $\hat{T}_d \: = \: \sum_{i, \sigma} \: [t_{d} \: d^{+}_{i+1, 
\sigma} \: d_{i, \sigma} \: + \: h.c \: ]$ with $d \: = \: c \: , \: f$
are hoppings for $c$ and $f$ electrons, \\
$\hat{V}_{0} \: = \: \sum_{i, \sigma} \: [ V_{0} \: f^{+}_{i, \sigma} \:
c_{i, \sigma} \: + \:  h.c \: ]$ is the on-site and  $\hat{V}_{1} \: =
\: \sum_{i, \sigma} \: [ \: V_{1} \: ( \: c^{+}_{i, \sigma} \: 
f_{i+1, \sigma} \: + \: f^{+}_{i, \sigma} \: 
c_{i+1, \sigma} \: ) \: + \: h.c. \: ]$ is the first-neighbor 
hybridization term, $\hat{U} \: = \: \sum_{i} \: n^{f}_{i, \uparrow} \:
n^{f}_{i, \downarrow}$ is the Hubbard on-site interaction, and 
$n^{d}_{i, \sigma} \: = \: d^{+}_{i, \sigma} \: 
d_{i, \sigma}$ are the particle-number operators. \\
To have a form of $H$ appropriate for our goal let us transform
Eq.(\ref{Hamiltonian}) to the following form:

\begin{eqnarray} \label{H1}
H \: &=& \: - \: \sum_{i, \sigma} \: ( \: |x|^2  \: + \: |y|^2 \: + \: |z|^2
\: + \: |v|^2 \: ) \: + \: \sum_{i, \sigma} \: a_{i, \sigma} \:  
a^{+}_{i, \sigma} \: +  \: ( \: |x|^2 \: + \: |y|^2 \: ) \: \sum_{i, \sigma} 
\: n^{c}_{i, \sigma} \: + \:
\nonumber \\
\quad && \: ( \: |z|^2 \: + \: |v|^2 \: + \: E_f \: ) \: \sum_{i, \sigma} \: 
n^{f}_{i, \sigma} \: + \: U \: \sum_{i, \sigma} \: n^{f}_{i, \uparrow} \: 
n^{f}_{i, \downarrow} 
\end{eqnarray}
where $a^{+}_{i, \sigma} \: = \: x \: c^{+}_{i, \sigma} \: + \: y \:  
c^{+}_{i+1, \sigma} \: + \:  z \:  f^{+}_{i, \sigma} \: + \: v \: 
f^{+}_{i+1, \sigma}$ and $x, y, z$ and $v$ are given by the following 
equations:
\begin{eqnarray} \label{xyzveqs}
x \: y^* \: = \: -\: t_c \: , \: \quad \: z \: v^* \: = \: - \: t_f \: ,
 \: \quad \: x \: z^* \: + \: y \: v^* \: = \: - \: V_0 \: ,  
 \: \quad \: x \: v^* \: = \: z \: y^* \: = \: - \: V_1.
\end{eqnarray}
In order to solve Eqs.(\ref{xyzveqs}) let us introduce $m=\frac{t_f}{V1}$. 
Using $m$, from Eqs.(\ref{xyzveqs}) we have  
$z \: = \: m \: x$, $v \: = \: m^* \: y$ and the square of $x$ is determined
by 
$|x|^2 \: = \: - \: \frac{V_0}{2 m^*} \: \pm \: \sqrt{ \: \frac{V_0^2}{4 
m^{*2}} \: - t_c^2  \: \frac{m}{m^*}}$. During the calculation we used the 
conditions 
\begin{eqnarray} \label{conditions}
t_f \:  t_c \: = \: V_1^2 \: , \: \quad \: 
\frac{V_0^2}{4 m^{*2}} \: + \:  t_c^2 \: \geq \: 0 \: , \: \quad \: - \: 
\frac{V_0}{2 m^*} \: \pm \: \sqrt{\frac{V_0^2}{4 m^{*2}} - t_c^2 \: 
\frac{m}{m^*}} \: \geq 0.
\end{eqnarray}
Going on with transforming  $H$ we get from Eq.(\ref{H1})
\begin{eqnarray} \label{H2}
H \: &=& \: P \: + \: P^{'} \: + \:  ( \: |x|^2  \: + \: |y|^2 \: ) \: 
\sum_{i, \sigma} \: n^{c}_{i, \sigma} \: + \: 
( \: |z|^2  \: + \: |v|^2  \: + \: E_f \: + \: U \: ) \: \sum_{i, \sigma} \: 
n^{f}_{i, \sigma} 
\nonumber \\
&+& \: K \:  - \:  U \:  L,  
\end{eqnarray}
where $L$ is the lattice size, $P \: = \: \sum_{i, \sigma} \: a_{i, \sigma} 
\: a^{+}_{i, \sigma} \: $, $ \: P^{'} \: = \: U \: \sum_i \: ( \: 1 \: - \:  
n^f_{i, \uparrow} \: - \:  n^f_{i, \downarrow} \: + \: n^f_{i, \uparrow} \: 
n^f_{i, \downarrow} \: )$ and $K \: = \: - \: \sum_{i, \sigma} \:  ( \: |x|^2
\: + \: |y|^2  \: ) \: ( \: |m|^2  \: + \: 1 \: )$.
After this step, from Eq.(\ref{H2}) in the case of  
$E_f \: = \:  ( \: 1 \: - \: |m|^2 \: ) \: ( \: |x|^2 \: + \: |y|^2 \: ) \: 
- \: U$, the final form of $H$ is given by 
\begin{eqnarray} \label{H_decomposed}
H \: = \: P \: + \:  P{'} \: + \: ( \: |x|^2 \: + \: |y|^2 \: ) \: 
\sum_{i, \sigma} \: ( \: n^c_{i, \sigma} \:  + \: n^f_{i, \sigma} \: ) \:  
+ \:  K \:  - \:  U \: L.
\end{eqnarray}
This form of $H$ consists of constants $K$ and $U L$, positive semidefinite 
operators $P$ and $P^{'}$, and $ \sum_{i, \sigma} \:  ( \: n^c_{i, \sigma} \:
+ \: n^f_{i, \sigma} \: )$ which is related to the particle number as
$\langle \: n \: \rangle  \: = \: 1/L \: \sum_{i, \sigma} \: 
( \: n^c_{i, \sigma} \: + \: n^f_{i, \sigma} \: )$. Because $P$ and $P^{'}$ 
are positive semidefinite, omitting them the ground state of the operator
built up of the remaining terms must be a lower bound to the ground-state 
energy:
\begin{eqnarray}\label{lowerbound}
E_l \: = \: ( \: |x|^2 \: + \: |y|^2 \: ) \: \langle \: n \: \rangle \: L \: 
+ \: K \: - \: U \: L.
\end{eqnarray}
An upper limit $E_u$ for the ground state energy $E_0$ can be derived from 
the variational principle, assuming some wave function $| \: \psi \: \rangle$
to be the ground state. If there is any domain in the space of the model 
parameters with $E_l \: = \: E_u \: = \: E_0$ then on that domain the exact 
ground state is $| \: \psi \: \rangle$ and the exact ground state energy is 
$E_0$. This procedure will be followed in the next Section. 

\section{The obtained solutions}

In this Section we describe two different and exact ground-state solutions for
PAM at finite $U$ related to an insulating and a metallic phase. We start the
presentation with the insulating case.

\subsection{The insulating state}

Consider that all the coupling constants are real. We have in this case
from Eqs.(\ref{xyzveqs})

\begin{eqnarray} \label{xyzv1}
x \: &=& \:  ( \: - \: \frac{V_0 V_1}{2 t_f} \: \pm \: \sqrt{\frac{V_0^2 \: 
V_1^2}{4 \: t_f^2} \: - \: t_c^2} \:)^{1/2} \: e^{i \: \phi_{x}} \: , 
\\ 
y \: &=& \: - \: t_c \: ( \: - \: \frac{V_0 \: V_1}{2 \: t_f} \: \pm \: 
\sqrt{\frac{V_0^2 \: V_1^2}{4 \: t_f^2} - t_c^2} \: )^{-1/2} \:  
e^{-i \phi_{x}} \: , \:  
\nonumber \\
z \: &=& \: \frac{V_1}{t_c} \: ( \: - \: \frac{V_0 \: V_1}{2 \: t_f} \: 
\pm \: \sqrt{\frac{V_0^2 \: V_1^2}{4 \: t_f^2} \: - \: t_c^2} \: )^{1/2} \:  
e^{i \: \phi_{x}} \: ,
\nonumber \\
v \: &=& \: - \: V_1 \: ( \: - \: \frac{V_0 \: V_1}{2 \: t_f} \: \pm \: 
\sqrt{\frac{V_0^2 \: V_1^2}{4 \: t_f^2} \: - \: t_c^2} \: )^{-1/2} \: 
e^{-i \: \phi_{x}},
\nonumber
\end{eqnarray}
where $\phi_{x}$ is an arbitrary phase. Inserting $x$ and $y$ into the 
expression of $E_f$ given above we have  
$E_f \: = \: - \: ( \: 1 \: - \: \frac{t_f^2}{V_1^2} \: ) \: 
\frac{V_0 V_1}{t_f} \: - \: U$.
We also need the conditions $ t_f  \: t_c \: = \: V_1^2 \:$, $\: 
\frac{V_0^2}{V_1^2} \: \geq \: 4 \: $, $ \: - \: \frac{V_0 V_1}{2 t_f} \: 
\geq \: 0$, that are counterparts of those given in Eq.(\ref{conditions}) 
but for the case of real coupling constants. In these conditions we have a 
lower limit for the ground state energy from Eq.(\ref{lowerbound}) and 
Eq.(\ref{xyzv1}), in the case of $\langle \: n \: \rangle \: = \: 3$:
\begin{eqnarray} \label{lowerlimit1}
E_{l, 1} \: = \: [ \: - \: \frac{V_0 V_1}{t_f} \: ( \: 1 \: - \:  2 \: 
\frac{t_f^2}{V_1^2} \: ) \: - \:  U \: ] \: L.
\end{eqnarray}
To get also an upper bound we call the variational principle. Let the trial 
wave function be
\begin{eqnarray} \label{psi1}
&&| \:  \psi_{1} \: \rangle \: =  \: \prod_{i} \:  ( \: c_{i, \uparrow}^{+} 
\: + \: m \: f_{i,\uparrow}^{+} \: ) \:  ( \:  c_{i, \downarrow}^{+} \: + 
\: m \: f_{i, \downarrow}^{+} \: ) \: ( \: \alpha_i \: c_{i, \uparrow}^{+} \:
+ \: \beta_i \:  c_{i, \downarrow}^{+}  \: + \: \gamma_i \: 
f_{i,\uparrow}^{+} \: + \: \delta_i \: f_{i, \downarrow}^{+} \: ) \: 
| \: 0 \: \rangle 
\\ 
&& = \: \prod_{i} \:   [ \:  \epsilon_i \: ( \: c_{i, \uparrow}^{+} \: c_{i,
\downarrow}^{+}  \: f_{i, \uparrow}^{+} \: - \: m \: f_{i, \uparrow}^{+} \: 
f_{i, \downarrow}^{+} \: c_{i, \uparrow}^{+} \: ) \: + \: \nu_i \:  ( \: 
c_{i, \uparrow}^{+} \: c_{i, \downarrow}^{+}  \: f_{i, \downarrow}^{+} \: 
- \: m \: f_{i, \uparrow}^{+} \: f_{i, \downarrow}^{+} \: 
c_{i, \downarrow}^{+} \: ) \: ] \: | \:  0  \: \rangle 
\nonumber
\end{eqnarray}
where $\epsilon_i$ and $\nu_i$ are related to the quantities 
$\alpha_i \: , \: \beta_i \: , \: \gamma_i \: , \: \delta_i$. Since 
$| \: \psi_{1} \: \rangle$ sets  three electrons on any lattice
site it belongs to band filling $\langle \: n \: \rangle \: = \: 3$. 
Using Eq.(\ref{psi1}) we have an upper limit for the ground state energy:
$E_{u, 1} \: = \: \langle \:  \psi_{1} \: | \: H \: | \: \psi_{1} \: 
\rangle \: /  \: \langle \:  \psi_{1} \: | \: \psi_{1} \: \rangle$. 
Since $P \: | \: \psi_{1} \: \rangle \: = \: P^{'} \: | \: \psi_{1} \: 
\rangle \: = \: 0,$ it follows that  $E_{u, 1} \: = \: 
E_{l, 1} \: = \: E_0$, where $E_{0}$ is the ground state energy. 
Collecting all the conditions we used:
$U \: \geq \: 0$, $t_f \: t_c \: = \: V_1^2$, $\frac{V_0^2}{V_1^2} \: 
\geq \: 4$, $ \: - \: \frac{V_0 V_1}{2 t_f} \: \geq \: 0, \quad
E_f \: = \: - \: ( \: 1 \: - \:  \frac{t_f^2}{V_1^2} \: ) \:  
\frac{V_0 V_1}{t_f} \: - \: U, \quad
\langle \: n \:\rangle \: = \: 3,
$
the ground state is given by Eq.(\ref{psi1}) and the 
ground state energy is of the form of Eq.(\ref{lowerlimit1}).
\\
To get a deeper insight into the nature of the state represented by
$| \: \psi_{1} \: \rangle$ we show some expectation values calculated with
 it. Since $| \: \psi_{1} \: \rangle$ sets three electrons on each site it is 
obvious that any way of moving an electron from a certain site $i$ to an 
other $j \: ( \: \neq \: i \: )$ results an orthogonal state to it so we have 
\begin{eqnarray} \label{insulatecond}
\langle \: T_c \: \rangle \: = \: \langle \: T_f  \: \rangle \: = 
\: \langle \: V_1 \: \rangle \: = \: 0 \: .
\end{eqnarray}
Expectation values of other terms after normalization are given as
\begin{eqnarray} \label{otherterms1}
\langle \: n_{i, \uparrow}^{f} \: \rangle \: &=& \: \frac{m^2}{1+m^2} \: + 
\: \frac{\epsilon_i^2}{(\epsilon_i^2 \: + \: \nu_i^2 \: ) \: ( \: 1 \: + 
\: m^2 \: )} \: ,   
\nonumber \\
\langle \: n_{i, \downarrow}^{f} \: \rangle \: &=& \: \frac{m^2}{1+m^2} \: 
+ \: \frac{\nu_i^2}{( \: \epsilon_i^2 \: + \: \nu_i^2 \: ) \: ( \: 1 \: + 
\: m^2 \: )} \: , 
\nonumber \\
\langle \: n_{i, \uparrow}^{c} \: \rangle  \: &= & \: \frac{1}{1+m^2} \: + \: 
\frac{\epsilon_i^2 m^2}{(\: \epsilon_i^2 \: + \: \nu_i^2 \: ) \: ( \: 1 \: + 
\: m^2 \: )} \: ,   
\nonumber \\
\langle \: n_{i, \downarrow}^{c} \: \rangle \: &=& \: \frac{1}{1+m^2} \: + \: 
\frac{\nu_i^2 m^2}{( \: \epsilon_i^2  \: + \: \nu_i^2 \: ) \: ( \: 1 \: + 
\: m^2 \: )} \: ,   
\nonumber \\
\langle \: n_{i, \uparrow}^{f} \: - \:  n_{i, \downarrow}^{f} \: \rangle \: 
&=& \: \frac{\epsilon_i^2 - \nu_i^2}
{(\epsilon_i^2  \: + \: \nu_i^2 \: ) \: ( \: 1 \: + \: m^2 \: )} \: ,  
\nonumber \\
\langle \: n_{i, \uparrow}^{c} \: - \:  n_{i, \downarrow}^{c} \: \rangle \: 
&=& \: \frac{m^2 \:  ( \: \epsilon_i^2  \: - \: \nu_i^2 \: )}{( \: 
\epsilon_i^2  \: + \: \nu_i^2 \: ) \: ( \: 1 \: + \: m^2 \: )} \: ,   
\nonumber \\
\langle  \: U \: \rangle \: &=& \: U \: \frac {m^2}{ \: 1 \: + \: m^2} \: 
L \: , \: \quad \langle  \: V_0  \: \rangle \: = \: V_0 \: 
\frac{2m}{ \: 1 \: + \: m^2} \: L \: .
\nonumber 
\end{eqnarray}
It is known that in the case of large $U$ the periodic Anderson model 
goes to the Kondo lattice model which has an interesting insulating 
ground state at half-filling called the Kondo-insulator. As can be seen, 
PAM in strong coupling limit is able to give an insulating phase not only
at half-filling, but also at 3/4 filling as presented here.
From Eq.(\ref{otherterms1}) we see that $| \: \psi_{1} \: \rangle$ contains 
all the possible spin directions (i.e. is degenerate in spin) and thus 
describes an insulating paramagnetic state with zero net magnetic moment. 
Figure 1. and 2. present the emergence domains of $|\: \psi_{1} \: \rangle$ 
in the parameter space. 

\subsection{The metallic state}

Let consider the parameters $t_c$ and $t_f$ real, $V_0$ and $V_1$ imaginary, 
the Hamiltonian remaining of course Hermitian. We have from
Eq.(\ref{xyzveqs})
\begin{eqnarray} \label{xyzv2}
x \: = \:  ( \: \frac{V_0 V_1}{2 t_f} \: + \: \sqrt{\frac{V_0^2 \: 
V_1^2}{4 \: t_f^2} \: + \:  t_c^2} \: )^{1/2} \: e^{i \phi_{x}},  
\\
y \: = \: - \: t_c \: ( \: \frac{V_0 V_1}{2 t_f} \: + \: \sqrt{\frac{V_0^2 \:
V_1^2}{4 \: t_f^2} \: + \:  t_c^2} \: )^{*-1/2} \: e^{-i \phi_{x}},
\nonumber \\
z \: = \: \frac{V_1}{t_c} \: ( \:  \frac{V_0 V_1}{2 t_f} \: + \: 
\sqrt{\frac{V_0^2 \: V_1^2}{4 \: t_f^2} \: + \: t_c^2} \: )^{1/2} \: 
e^{i \phi_{x}}, 
\nonumber \\
v \: = \:  - \:  V_1^* \: ( \: \frac{V_0}{2 m} \: + \: \sqrt{\frac{V_0^2}{4 \:
m^2} \: + \: t_c^2} \: )^{*-1/2} \: e^{-i \: \phi_{x}} ,
\nonumber
\end{eqnarray}
where $\phi_{x}$ is arbitrary. Similarly to the previous case, from $x$ and 
$y$ given here we have $E_f = -2 (1-\frac{t_f^2}{|V_1|^2}) 
\sqrt{ \frac{V_0^2
V_1^2}{4 t_f^2} + t_c^2}- U$. The only condition coming from 
Eq.(\ref{conditions}) is $t_f t_c =V_1^2$ because the others are 
automatically valid. In the case of $\langle \: n \: \rangle = 3$  from 
Eq.(\ref{lowerbound}) and Eq.(\ref{xyzv2}) we find a lower limit
for the ground state energy as
\begin{eqnarray} \label{lowerlimit2}
E_{l, 2} \: = \: [ \:  2 \:  \sqrt{ \frac{V_0^2}{4 m^2} + t_c^2} \:  
( \: 1 \: - \: 2 \: \frac{t_f^2}{|V_1|^2} \: ) \: - \: U \: ] \: L
\end{eqnarray}
The upper limit to $E_0$ is derived with the trial function 
\begin{eqnarray} \label{psi2}
| \: \psi_{2} \: \rangle \: = \: \prod_{i} \:  x_{i, \uparrow}^{+} \:
x_{i, \downarrow}^{+} \: ( \: \alpha_{i} \: f_{i, \uparrow}^{+} \: + \: 
\beta_{i} \: f_{i, \downarrow}^{+} \: ) \: | \: 0 \: \rangle
\\
\hbox{where} \: \quad \: \alpha_{i} \: , \: \, \: \beta_{i} \: \quad \: 
\hbox{arbitrary,} \: \quad \: 
x_{i, \sigma}^{+} \: = \: d_{i, \sigma}^{+} \: \, \: \hbox{or} \: \, 
\:  e_{i, \sigma}^{+}  \quad \: \forall \: i \: \quad \: \hbox{and} \: 
 \nonumber \\
\quad \: d_{i, \sigma}^{+} \: = \: x \: ( \: c_{i, \sigma}^{+}\: + 
\: m \: f_{i, \sigma}^{+} \: ) \: , \:  
\quad \: e_{i, \sigma}^{+} \: = \: y \: ( \: c_{i+1, \sigma}^{+} \: + 
\: m \: f_{i+1, \sigma}^{+} \: )
\nonumber 
\end{eqnarray}
and $x$, $y$ are given in Eqs.(\ref{xyzv2}). It can be verified that 
$| \: \psi_{2} \: \rangle$ belongs to $\langle \: n \: \rangle \: = \: 3$.
The upper limit to the ground state energy is
$E_{u, 2} \: = \: \langle \: \psi_{2} \: | \: H \: | \: \psi_{2} \: \rangle /
\: \langle \: \psi_{2} \: | \psi_{2} \: \rangle$. 
Since $P \: | \: \psi_{2} \: \rangle \: = \: P^{'} \: | \: \psi_{2} \: 
\rangle \: = \: 0,$ it follows that  $E_{u, 2} \: =
\: E_{l, 2} \: = \: E_0$ is the ground state energy. Collecting all conditions: 
$U \: \geq \:  0$, $t_f \: t_c \: = \: V_1^2 $,
$E_f \:  = \: - \: 2 \:  ( \: 1 \: - \: \frac{t_f^2}{|V_1|^2} \: ) \:  
\sqrt{ \frac{V_0^2 V_1^2}{4 t_f^2} + t_c^2} \: - \:  U$, 
$\langle n \rangle = 3$ we find that the ground state is given in
Eq.(\ref{psi2}) and the ground state energy is as shown in 
Eq.(\ref{lowerlimit2}). In order to understand the physical behavior of this
state let us examine expectation values calculated with it.
Imposing the constraint in Eq.(\ref{psi2}) $\alpha \: = \: \alpha_{i}$, 
$\beta \: = \: \beta_{i}$ $\forall \: i$
we can express $| \: \psi_{2} \: \rangle$ in the momentum space:
\begin{eqnarray}
| \: \psi_{2} \: \rangle \: &=& \: \prod_{j} \: [ \: x \: 
( \: c_{j, \uparrow}^{+} \: + \: m \: f_{j, \uparrow}^{+} \: ) \: + 
\: y \:  ( \: c_{j+1, \uparrow}^{+} \: + 
\: m^{*} f_{j+1, \uparrow}^{+} \: ) \:  ]  \:  \times
\nonumber \\
&& 
\: \times \: [ \: x \: ( \: c_{j, \downarrow}^{+} \: + 
\: m \: f_{j, \downarrow}^{+} \: ) \: + \:
y \:  ( \:  c_{j+1, \downarrow}^{+} \:  +  \:
m^{*} \:  f_{j+1, \downarrow}^{+} \:  ) \:  ] \:  
[ \: \alpha \: f_{j, \uparrow}^{+} \: + \:
\beta f_{j, \downarrow}^{+} \: ] \: | \: 0 \: \rangle
\nonumber \\
\: &=& \: \prod_{j} \: [ \: \sum_{k_1} \: d_{k_1, \uparrow}^{+} 
\: e^{-i k_1 j} \: ] \: [  \: \sum_{k_2} \: 
d_{k_2, \downarrow}^{+} e^{-i k_2 j}]
[ \: \sum_{k_3} \: ( \: \alpha \: f_{k_3, \uparrow}^{+} \: + \: 
\beta \: f_{k_3, \downarrow}^{+} \: ) \:  e^{-i k_3 j} \: ] \: 
|\:  0 \: \rangle \:  
\nonumber \\
&& \: \hbox{with} \: \quad \: d_{k_i, \sigma}^{+} \: = 
\: ( \: x \: + \: y \: e^{-i k_i} \: ) \: c_{k_i, \sigma}^{+} \: + \: 
( \: x \: m \: + \: y \: m^{*} \: e^{-i k_i} \: ) \:  f_{k_i, \sigma}^{+} 
\end{eqnarray}
From this expression we find
\begin{eqnarray}
| \: \psi_{2} \: \rangle \: &=& \: M \: \prod_{k} \: d_{k, \uparrow}^{+} 
\: d_{k, \downarrow}^{+} \: ( \: \alpha \: f_{k, \uparrow}^{+}
\: + \: \beta \: f_{k, \downarrow}^{+} \: ) \: | \: 0 \: \rangle 
\nonumber \\
\: &=& \: M \: \prod_{k} \: [ \: e_1^2 \: \alpha \: c_{k, \uparrow}^{+}  
\: c_{k, \downarrow}^{+} \:  f_{k, \uparrow}^{+}
\: + \:  e_1 \:  e_2  \: \alpha \: c_{k, \uparrow}^{+} \: f_{k, 
\downarrow}^{+} \: f_{k, \uparrow}^{+} \: + \:  
\nonumber \\
&+& \: e_1^2  \: \beta \: c_{k, \uparrow}^{+}  \: c_{k, \downarrow}^{+} \: 
f_{k, \downarrow}^{+}\: + \:  e_1 \: e_2 \: \beta \: f_{k, \uparrow}^{+} \: 
c_{k, \downarrow}^{+}  \:  f_{k, \downarrow}^{+} \: ] | \:  0 \rangle
\nonumber \\
&& \: e_1 \: = \: x \: + \:  y \: e^{-i k} \: , \: \quad \: e_1^2 \: = \: 
|x|^2  \: + \: |y|^2 \: + \: 2 \: t_c \: cos(k), 
\nonumber \\
&& \: e_2 \: = \: x \: m \: + \: y \: m^{*} \: e^{-i k} \: , \: \quad 
\: e_2^2 \: = \: |m|^2  \: ( \: |x|^2  \: + \: |y|^2 \: ) \: + \: 2 \: 
m^2 \: t_c \: cos(k) \: ,
\nonumber \\
&& \: M \: = \: \sum_{(P)} \: (-1)^P \: e^{-3 i [ 1 p(1) + 2 p(2) + 
\dots + N p(N)]}
\end{eqnarray}
where $P$ denotes a permutation 
$( \: p(1) \: , \:  p(2) \: \dots p(N) \:)$ of $( \: 1 \:  \dots \: N \: )$.

The norm of $| \psi_{2} \rangle$ is given as
\begin{eqnarray}
\langle \psi_2 | \psi_2 \rangle \: &=& \: | K|^2 \:  ( \: |\alpha|^2 \: + 
\: |\beta|^2 \: ) \: \prod_{k} \: ( \: |e_1|^4 \: + \: |e_1|^2 \: |e_2|^2 \: )
\nonumber \\
&=& \:| K|^2  \: ( \: |\alpha|^2 \: +\: |\beta|^2 \: ) \: \prod_{k}
\:  [ \: u^2 \: ( \: 1 \: + \: |m|^2 \: ) \: + \: 4 \: t_c \: u \: cos(k) \:
\nonumber \\
&+& \: 4 \: t_c  \: ( \: 1 \: - \: |m|^2 \: ) \:  cos^2 (k) \: ] \: ,
\nonumber
\end{eqnarray}
where $u \: = \: |x|^2 \: + \: |y|^2$. After this step normed expectation 
values of different quantities can be given: 
\begin{eqnarray} \label{metalcond}
\langle  n_{c, \uparrow}(k)   \rangle \: &=& 
\: \frac{  |\alpha|^2}{ \: |\alpha|^2 \: + \: |\beta|^2} 
\\ 
&+& \: \frac{|\beta|^2}{ \: |\alpha|^2 \: + \: |\beta|^2} \: 
\frac{u^2 \: + \: 4 \: t_c \: u \: cos(k) \: + \: 4 \: t_c^2 \: cos^2(k)}{
u^2 \:  ( \: 1 \: + \: |m|^2 \: )  \: + \: 4 \: t_c \: u \: cos(k) \: + 
\: 4 \: t_c \:  ( \: 1 \: - \: |m|^2 \: ) \: cos^2 (k)},
\nonumber  \\
\: \langle  n_{c, \downarrow}(k) \rangle \: &=& \: 
\frac{|\beta|^2}{|\alpha|^2 \: + \: |\beta|^2} 
\nonumber \\ 
&+& \: \frac{|\alpha|^2}{|\alpha|^2 \: + \: |\beta|^2} \: 
\frac{u^2 \: + \: 4 \: t_c \: u \: cos(k) \: + \: 4 \: t_c^2 \: cos^2(k)}{
u^2 \:  ( \: 1 \: + \: |m|^2 \: ) \: + \: 4 \: t_c \: u \: cos(k) \: + 
\: 4 \: t_c \: ( \: 1 \: - \: |m|^2 \: ) \: cos^2 (k)},
\nonumber \\
\: \langle  n_{f, \uparrow}(k) \rangle \: &=& \: 
\frac{|\alpha|^2}{|\alpha|^2 \: + \: |\beta|^2} 
\nonumber \\
&+& \: \frac{|\beta|^2}{|\alpha|^2 \: + \: |\beta|^2} \:
\frac{|m|^2  \: ( \: u^2 \: - \: 4 \: t_c^2 \: cos^2(k) \: )}{u^2 \:  
( \: 1 \: + \: |m|^2 \: ) \: + \: 4 \: t_c \: u \: cos(k) \: + 
\: 4 \: t_c \: ( \: 1 \: - \: |m|^2 \: ) \: cos^2 (k)},
\nonumber \\
\: \langle n_{f, \downarrow}(k)  \rangle \: &=& \: 
\frac{|\beta|^2}{|\alpha|^2 \: + \: |\beta|^2} 
\nonumber \\
&+& \: \frac{|\alpha|^2}{|\alpha|^2 \: + \: |\beta|^2} \: 
\frac{|m|^2 \:  ( \: u^2 \: - \: 4 \: t_c^2 \: cos^2(k) \: )}{u^2 \:
( \: 1 \: + \: |m|^2 \: ) \: + \: 4 \: t_c \: u \: cos(k) \: + 
\: 4 \: t_c \: ( \: 1 \: - \: |m|^2 \: ) \:  cos^2 (k)},
\nonumber \\
\: \langle f^{+}_{\uparrow}(k) c_{\uparrow}(k) \rangle \: &=& \:  
\frac{-|\beta|^2 \:  ( \: u \: + \: 2 \: t_c \: cos(k) \: ) 
\: ( \: V_0 \: + \: 2i \: V_1 \: sin(k))}
{u^2 \:  ( \: 1 \: + \: |m|^2 \: ) \: + \:  4 \:  t_c \: u \: cos(k) \: + 
\: 4 \: t_c \:  ( \: 1 \: - \: |m|^2 \: ) \: cos^2 (k)},
\nonumber \\
\: \langle f^{+}_{\downarrow}(k) c_{\downarrow}(k) \rangle \: &=& \: 
\frac{-|\alpha|^2 \:  ( \: u \: + \: 2 \: t_c \: cos(k) \: ) 
\: ( \: V_0 \: + \: 2i \: V_1 \: sin(k) \: )}
{u^2 \:  ( \: 1 \: + \: |m|^2 \: ) \:  + \: 4 \: t_c \: u \: cos(k) \: + 
\: 4 \: t_c \: ( \: 1 \: - \:|m|^2 \: ) \: cos^2 (k)}
\nonumber \\
\: \langle n^d_{j, \sigma} \rangle \: &=& 
\: \sum_{k} \: \langle n^d_{\sigma}(k) \rangle \quad \: \forall  \:j, 
\quad \: d={c,f} \: \quad 
\sigma \: =  \: {\uparrow, \downarrow}
\nonumber
\end{eqnarray}
From Eqs.(\ref{metalcond}) it is seen that $| \: \psi_{2} \: \rangle$ 
describes an insulating state. This state is also paramagnetic because
$| \: \psi_{2} \: \rangle$ is degenerate in spin. This degeneration is
given by the arbitrary quantities $\alpha$ and $\beta$.
Furthermore, because $\langle \: n_{k} \: \rangle \: = \: 
\langle \:  n_{c, \uparrow}(k) \: + \:  n_{c, \downarrow}(k) \: + 
\: n_{f, \uparrow}(k) \: + \: n_{f, \downarrow}(k) \: \rangle \: = \: 3$ 
is $k$ independent, in $\langle \: n_{k} \: \rangle $ there is no any change
present at the Fermi-momentum $k_F$. Fig 3. presents the 
$| \:  \psi_{2} \: \rangle$ phase in the parameter 
space while Fig 4.  shows the relative situation of the phases 
$| \: \psi_{1} \: \rangle$ and  $| \: \psi_{2} \: \rangle$.

\subsection{Comparison of the two states}

Let us compare the wave functions $| \: \psi_{1} \: \rangle$ and 
$| \: \psi_{2} \: \rangle$.  The mathematical relationship between 
Eq.(\ref{psi1}) and Eq.(\ref{psi2}) can be seen in more transparent form 
after an orthogonalisation procedure. For this reason we decompose 
Eq.(\ref{psi2}) as follows
\begin{eqnarray}
| \: \psi_{2}^{'} \: \rangle \: &=& \: | \: \psi_{2} \: \rangle \: - 
\: 2  \: \prod_{i} \: d_{i,  \uparrow}^{+} \: d_{i, \downarrow}^{+} \: 
( \: \alpha_i \: f_{i, \uparrow}^{+} \: + 
\:\beta_i  \:  f_{i, \downarrow}^{+} \: ) \: | \: 0 \: \rangle \:  
\nonumber \\
&=& \: | \: \psi_{2} \: \rangle \: - \: 2 \: \prod_{i} \: e_{i,
\uparrow}^{+} \: e_{i, \downarrow}^{+} 
\: ( \: \alpha_i \: f_{i, \uparrow}^{+} \: + 
\:  \beta_i \:  f_{i, \downarrow}^{+} \: ) \: | \: 0 \: \rangle
\end{eqnarray} 
The wave-vector $ \: | \: \psi_{2}^{'} \: \rangle$ is zero if $m$ is a real 
number. This is seen from the fact that $| \psi_{2}^{'} \rangle$ contains at 
least one term of the form
\begin{eqnarray}
e_{j, \sigma}^{+} \: d_{j+1, \sigma}^{+} \: = 
\: x \: y \: ( \: c_{j+1, \sigma}^{+} \: + \: m^{*} \: 
f_{j+1, \sigma}^{+} \: ) \:  ( \: c_{j+1, \sigma}^{+} \: + \: m \: 
f_{j+1, \sigma}^{+} \: ) 
\end{eqnarray} 
which, due to the fermionic anticommutation rules, is zero if $m$ is real. 
It is also seen that in the case of infinite lattice
$\prod_{i} \:  d_{i, \uparrow}^{+} \: d_{i, \downarrow}^{+} \:  
( \: \alpha_i \: f_{i, \uparrow}^{+} \: + \:  \beta_i  \: 
f_{i, \downarrow}^{+} \: ) \: |  0 \rangle \: = \: 
\prod_{i} \: e_{i,  \uparrow}^{+} \:  e_{i, \downarrow}^{+} \:
( \: \alpha_i \:  f_{i, \uparrow}^{+} \: + \beta_i \:  
f_{i, \downarrow}^{+} \: ) \: | 0 \rangle \: = \: | \psi_1 \rangle$.
The state $| \: \psi_1 \: \rangle$ sets rigorously 3 electrons on each site, 
while $| \:  \psi_2^{'} \: \rangle$ contains at least one site with 4 
electron an one with 2 electrons. As a consequence, $| \: \psi_1 \: \rangle$ 
and $| \: \psi_2^{'} \: \rangle$ are orthogonal.
If m is a real number then the ground-state is  $| \: \psi_1 \: \rangle$,
and, if m is imaginary then the ground-state is $| \: \psi_2  \: \rangle \: = 
\: | \: \psi_2^{'}\: \rangle \: + \: 2 \: | \: \psi_1 \: \rangle$. 
The fact that 
$ \langle  \: \psi_1 \: | \: T_{c} \: | \:  \psi_1 \: \rangle \: = \: 0$ but 
$ \langle \: \psi_2 \: | \: T_{c} \: |  \:  \psi_2  \: \rangle \: \neq \: 0$ 
is attributed to the effect of $| \: \psi_2^{'} \: \rangle$ in the ground
state.

\section{Conclusion}

Using a positive semidefinite decomposition of the Hamiltonian we have found 
exact results for the periodic Anderson model at finite $U$ in $d=1$ 
dimension and $3/4$ band filling. The result detailed in first is for the 
case when all the coupling parameters have real value. In this case we 
found that the ground state is Eq.(\ref{psi1}) and the ground state energy is
of the form of Eq.(\ref{lowerlimit1}). The ground state Eq.(\ref{psi1}) 
describes a paramagnetic insulating state that interestingly is present
also in the strong coupling limit (i.e. Kondo lattice case) at 
three-quarters band filling.

The second solution presented is for the case when the hopping amplitudes are 
real and the hybridization amplitudes are imaginary (the Hamiltonian remaining
Hermitian). We found a ground state presented in Eq.(\ref{psi2}) and a 
ground state energy as given in Eq.(\ref{lowerlimit2}). 
This state has nonzero expectation values for the $c$ and $f$ hoppings
so it is a metallic state. We have found that the state has no discontinuities in 
$n_{k}$ at $k_F$. 

\acknowledgements
Research supported by contract OTKA-022874 of Hungarian 
Founds for Scientific Research.

\newpage

\section*{Figure captions}

Fig.1.:
The $| \: \psi_{1} \: \rangle$ insulating state in the phase diagram. Notations are 
$X \: = \: \frac{V_1}{t_f}$,  $Y \: = \: \frac{V_0}{t_f}$,
$Z \: = \: \frac{U}{t_f},$  $\alpha \: = \: \frac{E_f}{t_f}$. Setting for example  
 $\alpha \: = \: - \: 0.5$ the quantum state 
 $| \: \psi_{1} \: \rangle$ is ground state when 
$Z \: = \: - \: ( \: 1 \: - \: \frac{1}{X^2}) \: X \:  Y \: - \: \alpha$, with
the constraints$- \: X \: Y \: > \: 0$, $\: \frac{Y^2}{X^2} \: \geq \: 4
$.

Fig.2.:
A part of the $| \: \psi_{1} \: \rangle$ phase surface. Notations are the same
as in Fig.1.

Fig.3.:
The $| \: \psi_{2} \: \rangle$ metallic state in the phase diagram. 
For this Figure the notations are 
$X \: = \: \frac{V_1^2}{t_f^2}$,  $Y \: = \: \frac{V_0^2}{t_f^2}$,
$Z \: = \: \frac{U}{t_f},$  $\alpha \: = \: \frac{E_f}{t_f}$. 
Coosing $\alpha \: = \: - \: 0.5$ quantum state  $| \: \psi_{2} \: \rangle$ 
 is found to be  ground state when 
$ Z \: = \: - \: ( \: 1 \: - \: \frac{1}{X} \: ) \: \sqrt{X^2 +
\frac{1}{4} X Y} \: - \: \alpha$.

Fig.4.:
The phases $| \: \psi_{1} \: \rangle$ and $| \: \psi_{2}  \: \rangle$  
in the phase diagram presented together.

\newpage
\thispagestyle{empty}

\begin{figure}[!htb]
\begin{center}
\epsfig{bbllx=0,bblly=0,bburx=450,bbury=250,file=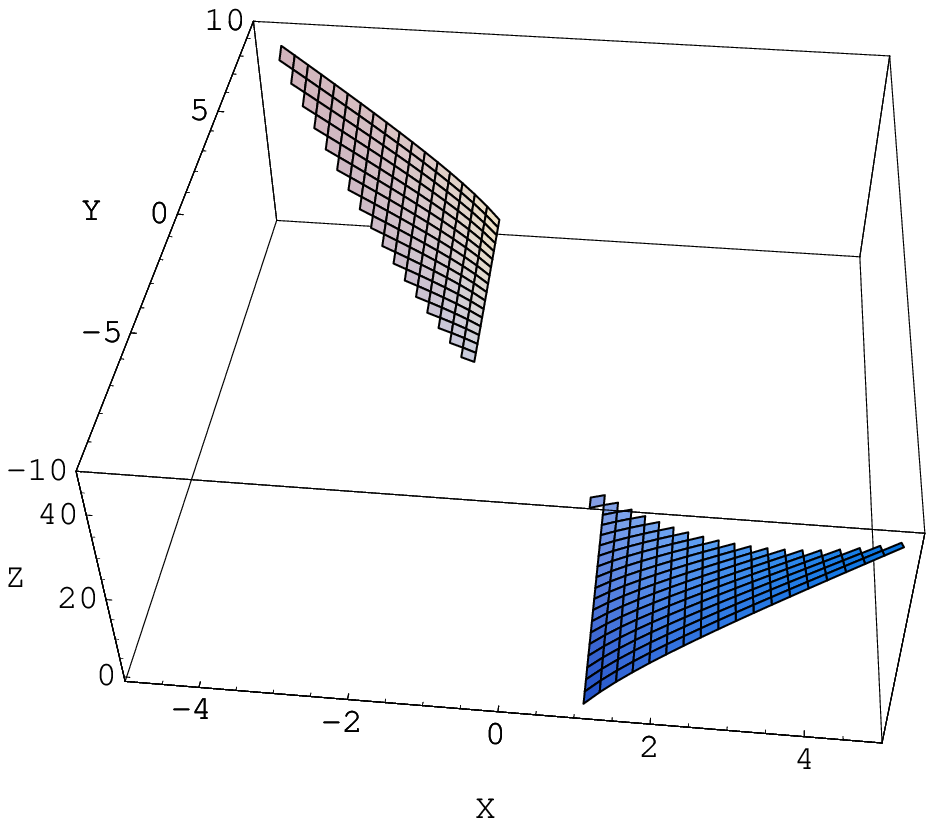}
\end{center} 
\end{figure}
\newpage
\thispagestyle{empty}

\begin{figure}[!htb]
\begin{center}
\epsfig{bbllx=0,bblly=0,bburx=450,bbury=250,file=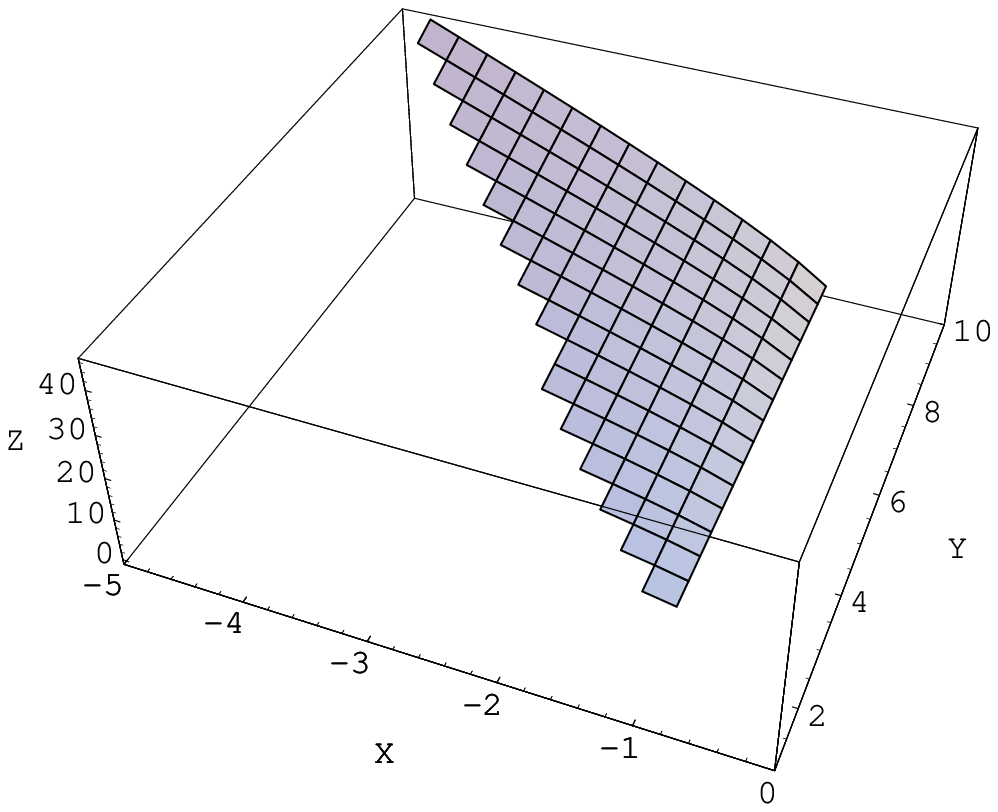}
\end{center} 
\end{figure}
\newpage
\thispagestyle{empty}

\begin{figure}[!htb]
\begin{center}
\epsfig{bbllx=0,bblly=0,bburx=450,bbury=250,file=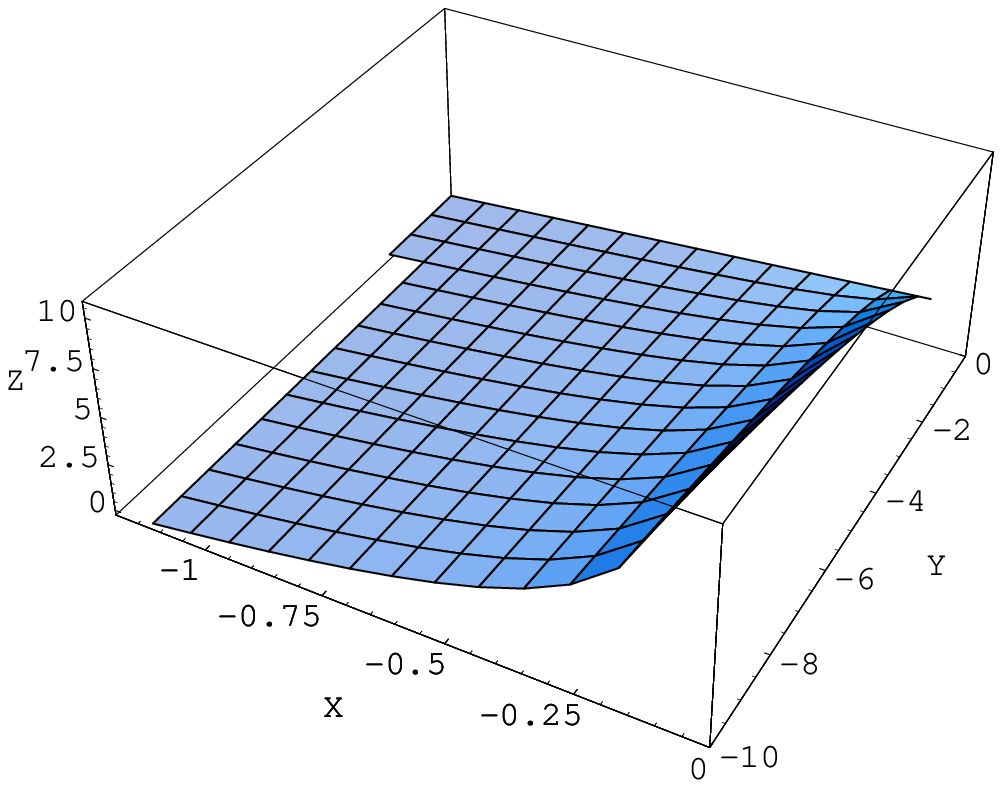}
\end{center} 
\end{figure}
\newpage
\thispagestyle{empty}

\begin{figure}[!htb]
\begin{center}
\epsfig{bbllx=120,bblly=360,bburx=610,bbury=800,file=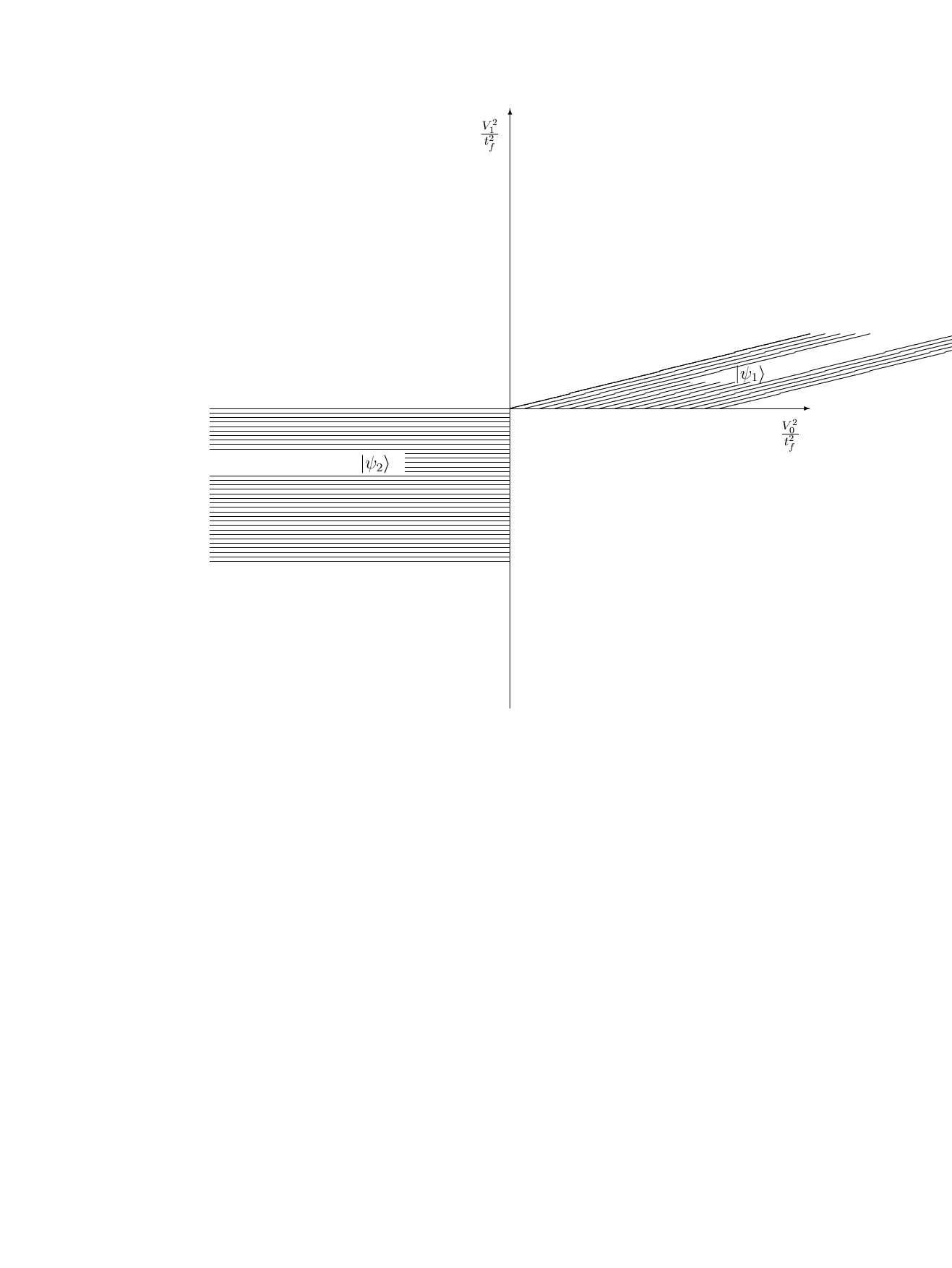, 
width=10cm, height=10cm}
\end{center} 
\end{figure}


\begin{thebibliography}{99}

\bibitem{b} U. Brandt, and A. Gieskus, {\sl Phys. Rev. Lett.} {\bf 68}, 
2648, (1992).  
\bibitem{s} R. Strack, {\sl Phys. Rev. Lett.} {\bf 70}, 833, (1993).    
\bibitem{o} I. Orlik, and Zs. Gul\'acsi, {\sl Phil. Mag. Lett.} 
{\bf 78}, 177, (1998). 

\end{thebibliography}
\end{document}